\documentclass[aps,pra,twocolumn,showpacs,groupedaddress,floatfix]{revtex4}

\usepackage{graphicx}
\usepackage{amsmath,amssymb}
\usepackage{times,txfonts}
\usepackage{color}      % use if color is used in text

\begin{document}

\title{Quasiparticle and quasihole states of nuclei around $^{56}$Ni}

\author{C. Barbieri}
\altaffiliation{Present address:Theoretical Nuclear Physics Laboratory, RIKEN Nishina Center, Japan.}
\affiliation{Theoretical Nuclear Physics Laboratory, RIKEN Nishina Center, 2-1 Hirosawa, Wako, Saitama 351-0198 Japan}
\affiliation{Gesellschaft f\"ur Schwerionenforschung Darmstadt, Planckstr. 1, D-64259 Darmstadt, Germany}

\author{M. Hjorth-Jensen}
\affiliation{Department of Physics and Center of Mathematics for Applications, University of Oslo, N-0316 Oslo, Norway}

\date{\today}

\begin{abstract}
 The single-particle spectral function of $^{56}$Ni has been computed within the framework of self-consistent Green's functions theory.  The Faddeev random phase approximation method and
 the G-matrix technique are used to account for the effects of long- and short-range physics on the spectral distribution.
 Large scale calculations have been performed in spaces including up to ten oscillator shells. The chiral N$^3$LO interaction is used together with a monopole correction that accounts for eventual 
missing three-nucleon forces.
 The single-particle energies associated with nucleon transfer to valence $1p0f$ orbits are found to be almost converged with respect to both the size of the model space and the oscillator frequency.
 The results support that $^{56}$Ni is a good doubly magic nucleus. The absolute spectroscopic factors
to the valence states on $A=55,57$ are also obtained.  For the transition between the ground states of $^{57}$Ni and $^{56}$Ni, the calculations nicely agree with heavy-ion knockout experiments.
\end{abstract}

\maketitle

%%%%%%%%%%%%%%%%%%%%%%%%%%%%%%%%%%%%%%%%%%%%%%%%%%%%%%%%%%%%%%%%%%%%%%%%%%%%%%%%%%%%%

\section{Introduction}
\label{intro}

The way shell closures and single-particle energies evolve as functions of the
number of nucleons is presently one of the greatest challenges
to our understanding of the basic features of nuclei. Doubly-magic nuclei are particularly important
and closed shell nuclei like $^{56}$Ni  and $^{100}$Sn have been the focus of several experiments
during the last years~\cite{ni56a,Yur.06,ni56c,ni56d,ni56e,ni56f,sn100a}. Their structure provides  
important information on theoretical interpretations  and our basic
understanding of matter.  In particular, recent experiments~\cite{ni56a,Yur.06,ni56c,ni56d,ni56e,ni56f} have aimed at extracting 
information about single-particle degrees of freedom in the vicinity of $^{56}$Ni. Experimental information
from single-nucleon transfer reactions  
and magnetic moments \cite{ni56a,Yur.06,ni56c,ni56d,ni56f}, can be used to extract and interpret complicated many-body wave 
functions in terms effective single-particle degrees of freedom. Transfer reactions provide for example
information about the angular distributions, the excitation energies and the spectroscopic factors of possible single-particles states. If one can infer from experimental data that
a single-particle picture is a viable starting point for interpreting a closed-shell nucleus
like $^{56}$Ni, one can use this nucleus as a basis for constructing valence-space 
effective interactions. These interactions  can in turn  be used
in shell-model calculations of nuclei with several valence nucleons above the $N=28$ and $Z=28$ filled shells of $^{56}$Ni. Recent measurements of spectroscopic factors of $^{57}$Ni in high-energy knockout reactions~\cite{Yur.06} seem to indicate that low-lying states in $^{57}$Ni can be characterized as single-particle states on top of $^{56}$Ni as a closed-core nucleus. Large-scale shell-model calculations by Horoi {\em et al}~\cite{horoi2006} corroborate these findings, whereas
a recent experiment on magnetic moments of the ground state of $^{57}$Cu \cite{ni56f}, expected to be described as 
one valence proton outside $^{56}$Ni, resulted 
in much smaller moments than those expected from a single-particle picture. 
Similarly, large transition matrix elements between the $0_1^+$ ground state
and the first excited $2_1^+$ state indicate that the $^{56}$Ni core is rather soft \cite{ni56e}, or stated differently, 
it implies a rather  fragmented single-particle picture.  
On the other hand, one ought keep in mind that quenchings of spectroscopic
factors to about 60\% are common even for good closed shell nuclei~\cite{lap93}.
Experimentally, spectroscopic factors are defined as the ratio
of the observed reaction rate with respect to the same rate calculated
assuming a full occupation of the relevant single-particle states.
They are therefore often interpreted as a measure of the occupancy
of a specific single-particle state. 
However, from a strict theoretical point of view spectroscopic factors 
are not occupation numbers but a measure of what fraction of
the full wave function can be factorized into a correlated state (often
chosen to be a given closed-shell core)  and an independent single-particle
or single-hole state. Large deviations from the values predicted by
an independent-particle model, point to a strongly correlated system.
In this regime, collective excitations that behave like single-particle
degrees of freedom---that is {\em quasiparticles}---can still arise.

The above mentioned large-scale shell-model calculations \cite{horoi2006}
have been performed in one major shell, the $1p0f$-shell,  with an effective
interaction fitted to reproduce properties of several nuclei that can be 
interpreted in terms of these single-particle states. The number
of possible Slater determinants that can be constructed when distributing 
eight valence protons and eight valence nucleons in the $1p0f$ shell is more than $10^9$.
This means that the inclusion of more complicated particle-hole excitations from 
shells below and above the 
$1p0f$ shell, are well beyond present capabilities
of large-scale diagonalization methods~\cite{Whitehead1977,caurier2005,dean2004b,navratil2004,horoi2006}.
The hope is that an effective interaction tailored to
one major shell includes as many as possible of these neglected particle-hole excitations.  
However, there are other many-body methods that allow for a 
computational scheme which accounts for a systematic
inclusion of more complicated many-body corrections.
Typical examples of such many-body methods are coupled-cluster methods~\cite{bartlett2007,helgaker,dean2004,Hag.08}, 
various types of Monte Carlo methods~\cite{pudliner1997,kdl97,utsuno1999}, 
perturbative many-body expansions~\cite{ellis1977,Hjo.95}, 
self-consistent Green's functions (SCGF) methods~\cite{DicVan,Dic.04,vanneck2002,vanneck2003,vanneck2005,Bar.01,Bar.02,Bar.03,Bar.06},
the density-matrix renormalization group~\cite{white1992,schollwock2005,dukelsky2002a,pittel2006} 
and {\em ab initio} density functional theory~\cite{bartlett2005,vanneck2006},
just to mention some of the available methods.

The Green's function Monte Carlo method
\cite{pudliner1997,pieper2001,wiringa2002,pieper2004} and the no-core
shell-model approach
\cite{navratil2004,navratil1998,navratil2000a,navratil2002,navratil2000b,navratil2003,nogga2006}
have been successfully applied to the theoretical description of
light nuclei with mass numbers $A \le 12$, and Hamiltonians based on
nucleon-nucleon and three-nucleon interactions.  However, present
experimental studies of nuclear stabilities are now being pushed to
larger mass regions, with mass numbers from $A=40$ to
$A=100$. Traditionally, this has been the realm of the nuclear shell-model
and nuclear density-functional theory. These methods employ Hamiltonians and density
functionals with phenomenological corrections and are not 
directly related to the vacuum nucleon-nucleon interaction employed in
{\em ab initio} calculations (exceptions are found when perturbative many-body methods are used \cite{Hjo.95}). However, in selected medium-mass nuclei, {\em ab initio} 
structure calculations can be performed using approaches like coupled-cluster and Green's functions theories.
These methods allow studying ground- and excited-state
properties of systems with dimensionalities beyond the capability of
present large-scale diagonalization approaches, with a much smaller
numerical effort when compared to diagonalization methods aiming at
similar accuracies. The accuracy of these methods is sufficiently high to
attribute an eventual disagreement between experimental data and
theoretical results to missing physics in the Hamiltonian. In this
way, such calculations help to increase our understanding
of the nuclear interaction on a very fundamental level.

Recent coupled-cluster calculations~\cite{Hag.08} 
have reported practically converged results of the ground 
state of medium-mass nuclei like $^{40}$Ca, $^{48}$Ni and $^{48}$Ca using the bare chiral  
interaction N$^3$LO~\cite{Ent.03}. 
These calculations employed a harmonic oscillator basis to construct the single-particle basis
and included correlations of the so-called singles and doubles types.  
It means that one-particle-one-hole
and two-particle-two-hole correlations acting 
on a many-body Slater determinant were summed to infinite order.
Recently, results with three-particle-three-hole correlations have also been obtained~\cite{hagen2009}.
The calculations were performed in a harmonic oscillator basis  containing up to fifteen major shells
and resulted in basically converged ground state properties for a given Hamiltonian.

Another method with a strong potential for performing {\em ab initio} calculations
of nuclei beyond $A = 12$ is self-consistent Green's functions theory.
Differently from the coupled-cluster and no-core shell model, this method
does not construct the wave function but evaluates directly the energies and
transition matrix elements for the transfer of one or more nucleons. 
Another important point is that the self-energy--the central component of the
formalism--has been shown to be an exact optical potential.
On the one hand, one can employ the formalism for pure {\em ab initio} studies. On the
other, the strong link with the response to experimental probes can be used to
constrain and improve phenomenological models.
 An example of this approach is the dispersive optical model recently derived
for chain of calcium isotopes~\cite{Cha.06,Cha.07}. This global optical potential
reproduces with high accuracy the known elastic scattering data, up to energies of 200~MeV.
Thus, Green's functions hold a promise of {\em both} bridging nuclear structure and 
reactions and for connecting the (relatively few) isotopes amenable of {\em ab initio}
calculations to the rest of the nuclear landscape.

In practical applications of Green's functions theory, one expands the self-energy in terms
of resummations of Feynman diagrams and truncates the series in a way that allows for
further systematic improvements of the formalism.
A powerful scheme for non-perturbative expansions is the so called Faddeev
random phase approximation (FRPA) that explicitly accounts for
particle-vibration couplings~\cite{Bar.01,Bar.07}. First applications of
this approach were devoted to $^{16}$O. One single calculation yielded the
basic information to be used for microscopic studies of spectroscopic
factors~\cite{Bar.02,Bar.09b}, the excitation spectrum~\cite{Bar.03},
two-nucleon emission~\cite{Bar.04,Mid.06} and the nucleon-nucleus
optical potential~\cite{Bar.05}.
Thus, the FRPA method pursues a global description of the many-body
dynamics, far beyond ground state properties alone.
Another difference with the coupled-cluster approach is that Green's functions
allow, via a diagrammatic approach, to directly introduce the correlations
outside the model space that are associated with short-range
degrees of freedom~\cite{Mut.95,Geu.96}. It can therefore be applied
to interactions with strong short-range cores (for example, the Argonne
model~\cite{Wir.95} was employed in Ref.~\cite{Bar.06}).

Self-consistent FRPA calculations of $^{16}$O were first performed
by the authors of Ref.~\cite{Bar.02}, and subsequently extended to fully
{\em ab initio} calculations in spaces up to eight oscillator shells~\cite{Bar.06}.
To our knowledge, the combination of the random phase approximation phonons and the proper treatment
of the energy dependence of the interaction vertex, makes this the most
accurate evaluation of single-particle states available for this nucleus.
In this work we extend the range of applications of this formalism  to studies of
  quasiparticle states around $^{56}$Ni, as a 
first application of Faddeev random phase approximation to $1p0f$ shell nuclei.
 As mentioned above, there is quite some experimental interest in single-particle
 properties around $^{56}$Ni.

This work is organized as follows.
Sec.~\ref{formalism} reviews the Faddeev random phase approximation approach and discusses the 
approximations made to calculate the self-consistent propagator.  
The convergence of single-particle properties is discussed in Sec.~\ref{calc} 
and the results for the spectral function are described in Sec.~\ref{res_spctfnct}.
 We refer the reader directly interested to discussion of physics results
to the latter section. Conclusions are drawn in Sec.~\ref{concl}.

\section{Formalism}
\label{formalism}

This section serves as an overview of the formalism we employ.
The treatment of short-range physics and the implementation
of self-consistency, which are improved with respect to previous works,
are discussed in details. The Faddeev random phase approximation
expansion is also introduced, but we refer the reader for more details 
in Refs.~\cite{Bar.01,Bar.07}.

In the framework of Green's function theory, the object of interest is the
single-particle propagator, instead of the many-body wave function.
In the following, greek indices $\alpha,\beta,\dots$ label the
orthonormal basis set of single-particle states included in the model space, while
latin indices $k,l, \dots$ refer to many-body states. We 
employ the convention of summing over repeated indices, unless specified
otherwise.
The single-particle propagator can be written in the so-called  Lehmann representation 
as~\cite{FetWal,DicVan}
\begin{equation}
 g_{\alpha \beta}(\omega) ~=~ 
 \sum_n  \frac{ \left( {\cal X}^{n}_{\alpha} \right)^* \;{\cal X}^{n}_{\beta} }
                       {\omega - \varepsilon^{+}_n + i \eta }  ~+~
 \sum_k \frac{ {\cal Y}^{k}_{\alpha} \; \left( {\cal Y}^{k}_{\beta} \right)^*  
}
                       {\omega - \varepsilon^{-}_k - i \eta } \; ,
\label{eq:g1}
\end{equation}
where ${\cal X}^{n}_{\alpha} = {\mbox{$\langle {\Psi^{A+1}_n} 
\vert $}} c^{\dag}_\alpha {\mbox{$\vert {\Psi^A_0} \rangle$}}$~
(${\cal Y}^{k}_{\alpha} = {\mbox{$\langle {\Psi^{A-1}_k} \vert $}} 
c_\alpha {\mbox{$\vert {\Psi^A_0} \rangle$}}$) are the spectroscopic 
amplitudes, $c_\alpha$~($c^\dag_\alpha$) are the second quantization 
annihilation (creation) operators and $\varepsilon^{+}_n = E^{A+1}_n -
 E^A_0$~($\varepsilon^{-}_k = E^A_0 - E^{A-1}_k$).
With these definitions, $\vert\Psi^{A+1}_n\rangle$ and $\vert\Psi^{A-1}_k\rangle$ 
are the eigenstates, while $E^{A+1}_n$ and $E^{A-1}_k$ are the corresponding energies of the 
($A\pm1$)-nucleon system. Therefore, the poles of the single-particle propagator reflect the 
energy transfer observed in pickup and knockout reactions.

The single-particle propagator $g_{\alpha \beta}(\omega)$ enters the Dyson equation as
\begin{equation}
 g_{\alpha \beta}(\omega) =  g^{0}_{\alpha \beta}(\omega) \; +  \;
 %\sum_{\gamma \delta}
 g^{0}_{\alpha \gamma}(\omega) 
     \Sigma^\star_{\gamma \delta}(\omega)   g_{\delta \beta}(\omega) \; .
\label{eq:Dys}
\end{equation}
It depends on the irreducible self-energy $\Sigma^\star(\omega)$. The 
latter can be written as the sum of two terms. The first terms describes the average
mean-field (MF) while the second term contains dynamic correlations, 
\begin{equation}
 \Sigma^{\star}_{\alpha \beta}(\omega) ~=~  \Sigma^{MF}_{\alpha \beta} 
   ~+~ \frac{1}{4} \, 
  %\sum_{\mu \nu \lambda \\ \gamma \delta \varepsilon}
  V_{\alpha \lambda , \mu \nu}
  ~ R_{\mu \nu \lambda , \gamma \delta \varepsilon}(\omega)
    ~ V_{\gamma \delta , \beta \varepsilon} \; .
\label{eq:Sigma1}
\end{equation}
In Eq.~(\ref{eq:Dys}), $g^{0}(\omega)$ is the so-called unperturbed single-propagator, 
corresponding to nucleons moving under the effect of the
kinetic energy part of the total Hamiltonian. The localization of the single-particle states
in the nuclear mean-field is due to the term $\Sigma^{MF}$, which extends
the Hartree-Fock potential to that of a fully correlated density matrix.
The term $V_{\alpha \beta , \gamma \delta}$ represents the antisymmetrized matrix 
elements of the nucleon-nucleon interaction. In this work, these will be approximated by an 
effective interaction, Eq.~(\ref{eq:Veff}), discussed in Sec.~\ref{form_bhf}.
Eq.~(\ref{eq:Sigma1}) introduces the two-particle-one-hole ({\em 2p1h}) and two-hole-one-particle
({\em 2h1p}) irreducible 
propagator $R(\omega)$. In its Feynman expansion, this propagator contains all diagrams
with any number of particle and hole lines {\em except} those that allow
the intermediate propagation of one single line. It can therefore be interpreted
as carrying the complete information on all configurations that cannot
be reduced to a nucleon interacting with the average nuclear field.
In particular, it includes the coupling of single-particle states to collective vibrations like
giant resonances.

The theoretical spectroscopic factors $Z_k$ and $Z_n$  for removal and addition, respectively, of
a nucleon, are given by the normalization integral of the corresponding
overlap wave functions. In the notation of Eq.~(\ref{eq:g1}), these are
\begin{eqnarray}
  Z_k &=&  \sum_\alpha   
 \left|  \langle {\Psi^{A-1}_k} \vert c_\alpha \vert {\Psi^A_0} \rangle  \right|^2
  =  \sum_\alpha\left| {\cal Y}^{k}_{\alpha} \right|^2  \; ,
\nonumber \\
  Z_n &=&  \sum_\alpha   
 \left|  \langle {\Psi^{A+1}_n} \vert c^{\dag}_\alpha \vert {\Psi^A_0} \rangle  \right|^2
  =  \sum_\alpha\left| {\cal X}^{n}_{\alpha} \right|^2   \; .
\label{def:Zkn}
\end{eqnarray}
The hole states are normalized according to
\begin{equation}
 Z_k = \sum_{\alpha}
\left| {\cal Y}^{k}_{\alpha} \right|^2
    =  1 + 
     \sum_{\alpha , \beta}   \left( {\cal Y}^{k}_{\alpha} \right)^* 
       \left.  \frac{ \partial \Sigma^*_{\alpha \beta}(\omega) }
                    { \partial \omega}
       \right|_{\omega = \varepsilon^{-}_k }
	       {\cal Y}^{k}_{\beta}  \; \; ,
\label{eq:sf}
\end{equation}
which follows directly from the Dyson equation~(\ref{eq:Dys}).
The same relation applies to particle states,
with~${\cal Y}^{k}_{\alpha}$ replaced by~$\left( {\cal X}^{n}_{\alpha} \right)^*$.
Because of the analytical properties of $\Sigma^\star(\omega)$, the derivative term
results always in a  negative contribution, leading thereby to a quenching
of the spectroscopic factors.

It must be stressed that Eqs.~(\ref{eq:Dys}) and~(\ref{eq:Sigma1}) do not
involve any approximation. Therefore, the full knowledge of  $\Sigma^{MF}$
and $R(\omega)$ would be equivalent to the exact solution of the Schr\"odinger
equation.
In practical calculations, it is always necessary to truncate the full 
space of available single-particle states  to a finite
model space and to select a limited set of many-body correlations.
 The approximations employed in the present work to evaluate these quantities
are discussed in the rest of this section.

\subsection{Short-range physics and effective interaction}
\label{form_bhf}

 The present calculations were performed within a large but finite set of
harmonic oscillator states, including single-particle states up to ten major shells.  
In order to treat the short-range part of the nucleon-nucleon interaction, one must
resum explicitly the series of ladder diagrams for two nucleons outside this model space.
These contributions are included in the self-energy in two different ways. Firstly, they are
explicitly added to the mean-field part, $\Sigma^{MF}$, in order to reconstruct the contribution
of short-range correlations in the full Hilbert space. Secondly, they are included in a
regularized effective Hamiltonian which is used to calculate the long-range part
of correlations---described by $R(\omega)$---inside the chosen model space.
This approach leads to calculating
the well known $G$-matrix~\cite{Hjo.95, CENS.08}, which is then used
as an energy dependent effective interaction inside
the model space~\footnote{We use a definition of $G(\omega)$ in which {\em all}
states belonging to the model space are Pauli blocked since diagrams among
them are already included in the many-body calculation of $R(\omega)$.}.
%
%In propagator theory, an energy dependent force can be
%treated correctly at the level of the mean-field.
 In this case the $\Sigma^{MF}$ part of the self-energy will also depend on energy. This is
given by
\begin{eqnarray}
%\begin{equation}
 \Sigma^{MF}_{\alpha \beta}(\omega) &=&
   i \sum_{\gamma \delta} \int \frac{d \omega '}{2 \pi} 
   G_{\alpha \gamma, \delta \beta}(\omega + \omega ')
      g_{\gamma \delta}(\omega ') 
\nonumber \\
&=& \sum_{\gamma \delta} \sum_{k} G_{\alpha \gamma, \beta \delta}(\omega +\varepsilon^{-}_k)
 \; {\cal Y}^{k}_{\delta} \; \left( {\cal Y}^{k}_{\gamma} \right)^*  \; ,
\label{eq:h_BHF}
%\end{equation}
\end{eqnarray}
where $G_{\alpha \beta , \gamma \delta}(\omega)$ are the matrix elements
of the $G$-matrix interaction. Eq.~(\ref{eq:h_BHF}) differs from the standard
Brueckner-Hartree-Fock %%(BHF)
potential in the fact that the mean-field
is not represented by a set of independent nucleons filling various orbits.
Rather, the medium is described by the hole spectroscopic amplitudes
${\cal Y}^{k}$ that lead to the fully correlated density matrix. Since the
latter are obtained by solving the Dyson equation, they must be obtained
iteratively in a self-consistent way. Whenever the second term
in Eq.~(\ref{eq:Sigma1}) is neglected, this procedure simply reduces to solving
the standard Brueckner-Hartree-Fock equations. However, as soon as the response of the medium
$R(\omega)$ is accounted for, the single-particle propagator becomes fragmented and
$\Sigma^{MF}$ describes the interaction of a particle with the ``correlated''
medium.

We remind that using the $G$-matrix in Eq.~(\ref{eq:h_BHF}) corresponds to summing
the mean-field term form Eq.~(\ref{eq:Sigma1}) and ladder diagrams with intermediate two-particle states 
outside
the chosen model space.
 This  partitioning procedure has two main consequences. First, the effects
of short-range physics at the two-body level on the total energy are included in the renormalized
interaction. This leads to a softer force that can be applied within a
``low-momentum'' model space. Secondly, due to the explicit energy dependence, the term
$\Sigma^{MF}(\omega)$ contributes as well to the normalization of spectroscopic
factors, Eq.~(\ref{eq:sf}). This provides a natural way to determine the amount
of strength that the free interaction would admix into configurations outside
the model space, see for example the discussion of Refs.~\cite{Mut.95,Geu.96}. 
Thus, the present approach differs from methods
based on the renormalization group, where instead an explicit renormalization of
the effective operators would be required~\cite{Ste.05}.

The energy dependence of the $G$-matrix becomes cumbersome in calculating the
polarization propagator. We define therefore a static effective interaction for
our model space to be used in our calculations of  the second term on the right-hand side
of Eq.~(\ref{eq:Sigma1}).  To do this we evaluate the average energy for the
{\em harmonic oscillator} single-particle states according to
\begin{equation}
 \varepsilon^{ho}_{\alpha} =
    \langle \alpha |  \frac{p^2}{2 m}  | \alpha \rangle ~+~ 
   \sum_{\beta \in F} 
    G_{\alpha  \beta , \alpha  \beta}(\omega = \varepsilon^{ho}_{\alpha} + \varepsilon^{ho}_{\beta})
 \; ,
 \label{ho_spe}
\end{equation}
where the sum is limited to those states that correspond to filled orbits
in the independent particle model.
Note that in Eq.~(\ref{ho_spe}) the single-particle energies are derived iteratively while
the oscillator wave functions remain unchanged.
Clearly, these orbits are
a crude approximation to the real quasiparticle states and will not
be used to construct $R(\omega)$ in Sec.~\ref{form_frpa}--which
will be rather determined in a self-consistent fashion.
The main purpose of the above procedure is to yield a prescription
for obtaining a starting energy independent effective interaction. Following
Gad and M\"uther~\cite{Gad.02}, we use the single-particle energies from
Eq.~(\ref{ho_spe}) to define an interaction for the given model space by
\begin{equation}
  V_{\alpha \beta , \gamma \delta}  = 
   \frac{1}{2} [
     G_{\alpha \beta , \gamma \delta}(\omega = \varepsilon^{ho}_{\alpha} + \varepsilon^{ho}_{\beta} )
 ~+~ G_{\alpha \beta , \gamma \delta}(\omega = \varepsilon^{ho}_{\gamma} + \varepsilon^{ho}_{\delta})
   ]
  \; .
 \label{eq:Veff}
\end{equation}
The $G$-matrix can be computed according Ref.~\cite{Hjo.95,CENS.08} for negative
energies, up to about $-5$ MeV. For larger values we fix the starting energy to
$\omega=-5$ MeV. Note that the starting energy
appearing in Eq.~(\ref{eq:h_BHF}) is shifted by the quasihole poles $\varepsilon^{-}_k$.
This ensures that the energy dependence of $\Sigma^{MF}(\omega)$ is fully accounted
for all  quasiparticle states in the $1p0f$ shell. 

%As a final remark, in the limit of very large model spaces the dependence
%on the starting energy becomes weak, while both $G(\omega)$
%and Eq.~(\ref{eq:Veff}) converge to the corresponding ``bare'' interaction. The usefulness 
%of the present approach relies on the separation of scales between long- and
%short-range physics. By prediagonalizing the latter (which corresponds to solving a two-particle problem) 
%it is then possible to 
%accelerate the convergence with respect a chosen the model space.  
%With a bare interaction many more shells would be needed in order to obtain an almost converged result. 

\subsection{The Faddeev random phase approximation method}
\label{form_frpa}

The polarization propagator $R(\omega)$ can be expanded in terms of simpler
Green's functions that involve the propagation of one or more quasiparticle states. 
This approach has the advantage that it aids in  identifying
key physics ingredients of the many-body dynamics. By truncating the expansion to a particular
subsets of diagrams or many-body correlations, one can then construct suitable approximations to
the self-energy. Moreover, since infinite sets of linked diagrams are summed
the approach is non-perturbative and satisfies the extensivity condition \cite{bartlett2007}. This
expansion also serves as a guideline for systematic improvements of the method.

\begin{figure}
\includegraphics[width=0.65\columnwidth,clip=true]{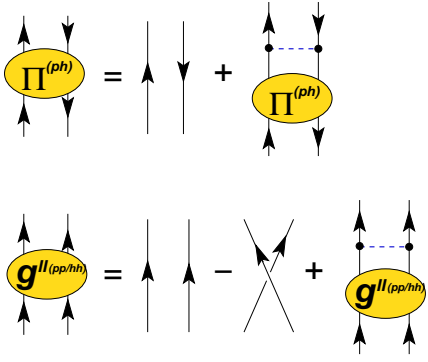}
\caption{(Color online) Diagrammatic equations for the polarization (above) 
and the two-particle (below) propagators in the RPA approach. Dashed lines 
are matrix elements of the effective nucleon-nucleon interaction, Eq.~(\ref{eq:Veff}).
 The full lines represent the independent-particle model propagator $g^{IPM}(\omega)$,
which is employed instead of the fully dressed one. See the text for details.}
\label{fig:rpaeq}
\end{figure}

\begin{figure}[t]
\includegraphics[width=0.7\columnwidth,clip=true]{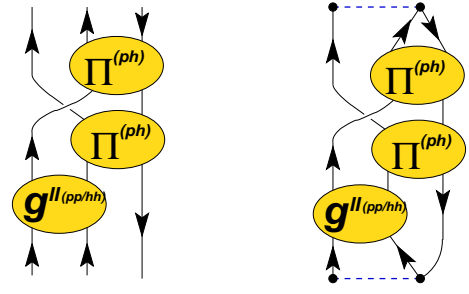}
\caption{(Color online)  Example of one of the diagrams that are summed to all 
orders by means of the Faddeev random phase approximation Eqs.~(\ref{eq:FaddTDA})~(left).
 The corresponding contribution to the self-energy, obtained upon insertion
 into Eq.~(\ref{eq:Sigma1}), is also shown~(right).}
\label{fig:faddex}
\end{figure}

Following Refs.~\cite{Bar.01,Bar.07}, we first consider the particle-hole
polarization propagator that describes excited states in the $A$-particle system 
\begin{eqnarray}
 \Pi_{\alpha \beta , \gamma \delta}(\omega) &=& 
%% g_{\alpha \beta}(\omega) ~=~ 
 \sum_{n \ne 0}  \frac{  {\mbox{$\langle {\Psi^N_0} \vert $}}
            c^{\dag}_\beta c_\alpha {\mbox{$\vert {\Psi^N_n} \rangle$}} \;
             {\mbox{$\langle {\Psi^N_n} \vert $}}
            c^{\dag}_\gamma c_\delta {\mbox{$\vert {\Psi^N_0} \rangle$}} }
            {\omega - \left( E^N_n - E^N_0 \right) + i \eta } 
\nonumber \\
 &-& \sum_{n \ne 0} \frac{  {\mbox{$\langle {\Psi^N_0} \vert $}}
              c^{\dag}_\gamma c_\delta {\mbox{$\vert {\Psi^N_n} \rangle$}} \;
                 {\mbox{$\langle {\Psi^N_n} \vert $}}
             c^{\dag}_\beta c_\alpha {\mbox{$\vert {\Psi^N_0} \rangle$}} }
            {\omega + \left( E^N_n - E^N_0 \right) - i \eta } \; ,
\label{eq:Pi}
\end{eqnarray}
and the two-particle propagator that describes the addition/removal of two 
particles
\begin{eqnarray}
 g^{II}_{\alpha \beta , \gamma \delta}(\omega) &=& 
%% g_{\alpha \beta}(\omega) &=& 
 \sum_n  \frac{  {\mbox{$\langle {\Psi^N_0} \vert $}}
                c_\beta c_\alpha {\mbox{$\vert {\Psi^{N+2}_n} \rangle$}} \;
                 {\mbox{$\langle {\Psi^{N+2}_n} \vert $}}
         c^{\dag}_\gamma c^{\dag}_\delta {\mbox{$\vert {\Psi^N_0} \rangle$}} }
            {\omega - \left( E^{N+2}_n - E^N_0 \right) + i \eta }
\nonumber \\  
&-& \sum_k  \frac{  {\mbox{$\langle {\Psi^N_0} \vert $}}
    c^{\dag}_\gamma c^{\dag}_\delta {\mbox{$\vert {\Psi^{N-2}_k} \rangle$}} \;
                 {\mbox{$\langle {\Psi^{N-2}_k} \vert $}}
                  c_\beta c_\alpha {\mbox{$\vert {\Psi^N_0} \rangle$}} }
            {\omega - \left( E^N_0 - E^{N-2}_k \right) - i \eta } \; .
\label{eq:g2}
\end{eqnarray}
%% We note that the expansion of $R(\omega)$ arises from applying the equations
%%of motion to the single-particle propagator~(\ref{eq:g1}), which is associated with the
%%ground state~$\vert\Psi^{A}_0\rangle$. Hence, all the Green's functions
%%appearing in this expansion will also be ground state based, including 
%%Eqs.~(\ref{eq:Pi}) and~(\ref{eq:g2}).
%% However, they contain in their Lehmann representations all the 
These Green's functions contain in their Lehmann representations all the 
relevant information regarding the excitation of particle-hole 
and two-particle or two-hole collective modes.
In this work we are interested in studying the influence of giant resonance
vibrations, which can be described within the random phase approximation (RPA).
In the Faddeev RPA approach, the propagators of Eqs.~(\ref{eq:Pi})
and~(\ref{eq:g2}) are then evaluated by solving the usual RPA equations,
which are depicted diagrammatically in Fig.~\ref{fig:rpaeq}.
 Since these equations reflect two-body correlations, they still have to be coupled
to an additional single-particle propagator, as in Fig.~\ref{fig:faddex}, to obtain
the corresponding approximation for the two-particle-one-hole and two-hole-one-particle components
of $R(\omega)$. This is achieved by solving two separate
sets of Faddeev equations, as discussed in Ref.~\cite{Bar.01}.

%%%%%%%%%%%%%%%%%%%%%%%%%%%%%%%%%%%%%%%%%%%%%%%%%%%%%%%%%%%%%%%%%%%%%%%%%%
%%%%%%%%%%%%%%%%%%%%%%%%%%%%%%%%%%%%%%%%%%%%%%%%%%%%%%%%%%%%%%%%%%%%%%%%%

 Taking the two-particle-one-hole (2p1h) case as an example, one can split $R^{(2p1h)}(\omega)$ 
in three different components $\bar{R}^{(i)}(\omega)$ ($i=1,2,3$) that differ from 
each other by the last pair of lines that interact in their diagrammatic 
expansion,
\begin{equation}
%\begin{eqnarray}
    \bar{R}^{(2p1h)}_{\alpha \beta \gamma , \mu \nu \lambda}(\omega) =
 \left[ {G^0}^>_{\alpha \beta \gamma , \mu \nu \lambda}(\omega)
        - {G^0}^>_{\beta \alpha \gamma ,  \mu \nu\lambda}(\omega) \right]
  + \sum_{i=1,2,3}     \bar{R}^{(i)}_{\alpha \beta \gamma , \mu \nu \lambda}(\omega)
  \; ,
\label{eq:faddfullR}
%\end{eqnarray}
\end{equation}
where ${G^0}^>(\omega)$ is the {\em 2p1h} propagator for three freely 
propagating lines.
These components are solutions of the following set of
Faddeev equations~\cite{Fad.61}
\begin{eqnarray}
  \lefteqn{
  \bar{R}^{(i)}_{\alpha  \beta  \gamma  ,
           \mu     \nu    \lambda   }(\omega) 
    ~=~ {G^0}^>_{\alpha  \beta  \gamma  ,
                 \mu'    \nu'   \lambda' }(\omega) ~
   \Gamma^{(i)}_{\mu'   \nu'     \lambda'  ,
                 \mu''  \nu''    \lambda'' }(\omega) }
        \hspace{.1in} & &
\nonumber  \\
  & \times & ~
  \left[ \bar{R}^{(j)}_{\mu''   \nu''  \lambda''  ,
                  \mu     \nu    \lambda    }(\omega) ~+~
         \bar{R}^{(k)}_{\mu''   \nu''  \lambda''  ,
                  \mu  \nu    \lambda    }(\omega)
  \right.
\label{eq:FaddTDA}  \\
  & & ~ ~+~ \left.
      {G^0}^>_{\mu''    \nu''   \lambda''   ,
               \mu      \nu     \lambda   }(\omega)
    - {G^0}^>_{\nu''    \mu''   \lambda''   ,
               \mu      \nu     \lambda   }(\omega)
       \right]  \; ,  ~ ~ i=1,2,3 ~ 
%       \right]  \;  ,
\nonumber
\end{eqnarray}
where ($i,j,k$) are cyclic permutations of ($1,2,3$).
The interaction vertices $\Gamma^{(i)}(\omega)$  contain the couplings of 
a particle-hole ({\em ph}), see Eq.~(\ref{eq:Pi}), or two-particle/two-hole ({\em pp/hh}), see Eq.~(\ref{eq:g2}), collective excitations
and a freely propagating line. 
The propagator $R(\omega)$ which we employ in Eq.~(\ref{eq:Sigma1}) is finally obtained by 
\begin{equation}
  R^{(2p1h)}_{\alpha  \beta  \gamma  ,
              \mu     \nu    \lambda   }(\omega)  ~=~
  U_{\alpha  \beta  \gamma , \mu'    \nu'   \lambda' } \;
   \bar{R}^{(2p1h)}_{\mu'   \nu'     \lambda'  ,
             \mu''  \nu''    \lambda''}(\omega)
    \;  U^\dag_{\mu''  \nu''    \lambda'' , \mu     \nu    \lambda} \; ,
\label{eq:URU}
\end{equation}
where the matrix $U$ has the effect of renormalizing the strength of
the dynamic self-energy. This correction ensures consistency with
perturbation theory up to third order.
  The explicit formulae of
the matrices $\Gamma^{(i)}(\omega)$ and~$U$ are given in terms of
the propagators of Eqs.~(\ref{eq:Pi}), (\ref{eq:g2}) and~(\ref{eq:gspref}), and the
interaction~$V_{\alpha \beta , \gamma \delta}$. They are discussed in detail
in  Ref.~\cite{Bar.07}.
The calculation of the {\em 2h1p} component of $R(\omega)$ follows completely analogous steps.
 
%%%%%%%%%%%%%%%%%%%%%%%%%%%%%%%%%%%%%%%%%%%%%%%%%%%%%%%%%%%%%%%%%%%%%%%%%
%%%%%%%%%%%%%%%%%%%%%%%%%%%%%%%%%%%%%%%%%%%%%%%%%%%%%%%%%%%%%%%%%%%%%%%%%

The present formalism includes the 
effects of {\em ph} and {\em pp/hh} motion simultaneously, 
while allowing interferences between these modes. 
These excitations
 are evaluated here at the RPA level and are then coupled to each other by 
solving Eqs.~(\ref{eq:FaddTDA}). This generates diagrams as the one displayed 
in Fig.~\ref{fig:faddex}. The Faddeev equations 
also ensure that the Pauli principle is correctly
taken into account at the {\em 2p1h} and {\em 2h1p} level. 
In addition, one can in principle employ dressed single-particle propagators
in these equations to generate a fully self-consistent solution, 
as done in Refs.~\cite{Bar.02,Bar.06} for valence orbits around~$^{16}$O.

\subsection{Self-consistent approach}
\label{form_selfcons}

In the self-consistent Green's function approach, both the $\Sigma^{MF}$
part of the self-energy and the polarization propagator $R(\omega)$
are expressed directly in terms of the exact single-particle propagator~$g(\omega)$.
The lines in Figs.~\ref{fig:rpaeq} and~\ref{fig:faddex} should thus represent
the fully dressed propagator obtained by solving  the Dyson equation.
Since the degrees of freedom contained in Eq.~(\ref{eq:g1}) are excitations
of the fully correlated system, the formalism does not depend on
an explicit reference state.
Normally, one first computes Eq.~(\ref{eq:Sigma1}) in terms of an
approximate propagator. The solution of Eq.~(\ref{eq:Dys}) is then used to
calculate an improved self-energy and the procedure is iterated to convergence.
Baym and Kadanoff have shown that the self-consistency requirement implies
the conservation of both microscopic and macroscopic properties~\cite{Bay.61,Bay.62}.
Intuitively, the self-consistency requirement becomes important whenever dynamical
correlations modify substantially the response with respect to the Hartree-Fock
mean-field (an example is the band-gap error problem in diamond crystals~\cite{Oni.02}).
 When applying standard Hartree-Fock theory to nuclear structure, most realistic
interactions predict unbound nuclei and valence orbits in the continuum.
This is a very poor starting point for any application of perturbation
theory and other many-body techniques. However, the self-consistent approach
requires using correlated quasiparticle energies and wave functions [the
poles and residues of Eq.~(\ref{eq:g1})]. These degrees of freedom form 
an optimal starting point for studies of  many-body dynamics at
the Fermi surface.

Accounting for the fragmentation of the single-particle propagator in the 
Faddeev random phase approximation
increases the computational load as one moves to larger nuclei and model spaces.
In this situation it is convenient to expand $R(\omega)$ in terms of an
independent-particle model  (IPM) 
propagator. This should approximate the dressed one but with a limited number of poles.
Thus, we  solve Eq.~(\ref{eq:FaddTDA}) in terms of
\begin{equation}
 g^{IPM}_{\alpha \beta}(\omega) ~=~ 
 \sum_{n   / \hspace{-0.12cm} \in F}  \frac{ \left( \phi^{n}_{\alpha} \right)^* \;\phi^{n}_{\beta} }
                       {\omega - \varepsilon^{IMP}_n + i \eta }  ~+~
 \sum_{k \in F} \frac{ \phi^{k}_{\alpha} \; \left( \phi^{k}_{\beta} \right)^*  
}
                       {\omega - \varepsilon^{IMP}_k - i \eta } \; ,
\label{eq:gspref}
\end{equation}
where $F$ represent the set of occupied orbits. The single-particle energies
$\varepsilon^{IPM}$ and wave functions $\phi$ are chosen such that 
$g^{IPM}(\omega)$ coincides with the real propagator $g(\omega)$
at the Fermi surface. To do this we define the following moments of
the poles of Eq.~(\ref{eq:g1}).
\begin{equation}
  M^p_{\alpha \beta} =
 \sum_n  \frac{ \left( {\cal X}^{n}_{\alpha} \right)^* \;{\cal X}^{n}_{\beta} }
                       {\left[ E_F - \varepsilon^{+}_n \right]^p}  +
 \sum_k \frac{ {\cal Y}^{k}_{\alpha} \; \left( {\cal Y}^{k}_{\beta} \right)^* }
                       {\left[ E_F - \varepsilon^{-}_k \right]^p} \; , ~ ~  p=0,1,2,\ldots
\label{eq:Mi}
\end{equation}
where $E_F$ is the Fermi energy. % and similarly for $g^{IPM}(\omega)$.
Eq.~(\ref{eq:gspref}) is determined by imposing
$M^{0,IPM}_{\alpha \beta}=M^0_{\alpha \beta}$ and
$M^{1,IPM}_{\alpha \beta}=M^1_{\alpha \beta}$.
The purpose of Eq.~(\ref{eq:Mi}) is to define a set of effective
single-particle orbits and energies that conserve the total spectroscopic strength
carried by the self-consistent propagator and the centroids of its
fragmented states.
While effective single-particle properties  form an appropriate starting point to evaluate $R(\omega)$,
it remains clear that they only represent average quantities.
Instead, it is Eq.~(\ref{eq:g1}) that must be related to experiment.

The propagator  $g^{IPM}(\omega)$ is derived from Eq.~(\ref{eq:g1}) and
it still needs to be evaluated in a iterative way. Therefore, the resulting
propagator $R(\omega)$  is (partially) self-consistent.
 We stress that Eq.~(\ref{eq:h_BHF}) can be calculated easily
from the fully dressed propagator. Thus self-consistency is achieved 
{\em exactly} at the mean-field level.

\section{Calculations and convergence}
\label{calc}

The present calculations were performed using a harmonic oscillator basis and
including up to ten major harmonic oscillator shells. We label these spaces 
with $N_{max}=3$, 5, 7, or 9, where $N=2n+l$. For the largest model space 
employed, $N_{max}=9$, all partial waves with orbital angular momentum $l\leq 7$ were
included. %Counting the third component of the total angular momentum,
This amount to 368 single-particle states for each particle species, protons and neutrons in our case.
The total number of available Slater determinants for $^{56}$Ni without any particular restrictions
is proportional to the product of the two binomials
\[
\left(\begin{array}{c} 368 \\ 28 \end{array}\right ) \times \left(\begin{array}{c} 368 \\ 28 \end{array}\right ),
\]
a number which clearly exceeds the capabilities of any direct large-scale diagonalization procedure.

The codes utilize a $jj$-coupling scheme to decouple the Faddeev equations of Eq.~(\ref{eq:FaddTDA}).
 At each iteration, the RPA equations are solved in the particle-hole and the particle-particle and hole-hole channels
using the single-particle orbits and energies from Eq.~(\ref{eq:gspref}). The resulting
propagators are inserted in Eqs.~(\ref{eq:FaddTDA}), which in turn are casted into a
non Hermitian eigenvalue problem~\cite{Bar.01}.  In our largest calculation
we diagonalize dense matrices of dimension up to $26\times 10^3$ two-particle-one-hole states. 
This number is
bound to increase when larger nuclei are investigated or more details of
nuclear fragmentation (that is more poles) are included
in $g^{IPM}(\omega)$~(\ref{eq:gspref}).
The numerical implementation of the Faddeev random phase approximation required
careful optimization
in evaluating the elements of the Faddeev matrix and a proper generalization
of the Arnoldi algorithm~\cite{anrref} to employ multiple pivots.
Similar improvements allowed to extend large scale calculations from the $A=16$ mass region 
to the $A=56$ mass region.
% A new FRPA code in C++ was developed for this purpose~\cite{barbieri_unpublished}. 
The dimensions reached in this work represent roughly the upper
limit when using table top single processor computers.
 Obviously, there is much to gain by taking advantage of modern supercomputer
facilities and future research efforts should be put into parallelization
of the present algorithms.

 Model spaces of eight to ten major shells are large enough for a proper
description of the response due to long-range correlations. These include
excitations of several MeVs into the region of giant resonances.
 The effects of short-range physics are also included by using a
$G$-matrix and an effective interaction as discussed in Sec.~\ref{form_bhf}.
These are derived using the  chiral nucleon-nucleon interaction N$^3$LO 
by Entem and Machleidt~\cite{Ent.03}. This interaction employs 
a cutoff of $\Lambda=500$ MeV.

Typical realistic two-nucleon interactions fail in reproducing the spin-orbit 
splittings and gaps between different shells. In particular, 
for the $N=28$ and $Z=28$ subshell closures these lead to an underestimation of the gap at the 
Fermi surface. In these cases, a complete diagonalization of the Hamiltonian 
would predict a deformed ground state of these nuclei even when they are 
experimentally known to be good spherical closed-shells systems~\cite{caurier2005,Hor.07}.
 This issue can be cured with a simple modification of the monopole strengths of the interaction. 
Recently, Zuker has reported that the same correction works well for several 
isotopes throughout the nuclear chart and proposed that this may be interpreted as  
a signature of missing three-nucleon interactions~\cite{Zuk.03}.
 
 The inclusion of three-nucleon interactions to the Faddeev random phase approximation formalism 
is beyond the scope of the present work. 
However, it will be shown in Sec.~\ref{calc_kdep} that properly reproducing
the Fermi gap is crucial in order to obtain meaningful results for the valence
space spectroscopic factors. Thus, we follow Ref.~\cite{Zuk.03} and modify the
monopoles in the N$^3$LO interaction model as
\begin{eqnarray}
 \Delta V^T_{fr} &\rightarrow&  \Delta V^T_{fr} - (-1)^T      \kappa_M \; ,
\nonumber \\
 \Delta V^T_{ff} &\rightarrow&  \Delta V^T_{ff} - 1.5 (1 - T) \kappa_M \; ,
\label{eq:monopole}
\end{eqnarray}
where $f$ and $r$ stand for the Brueckner-Hartree-Fock states associated to the $0f_{7/2}$ and
the ($1p_{3/2}$,$1p_{1/2}$,$0f_{5/2}$) orbits, respectively. In the limit
of large spaces, the Brueckner-Hartree-Fock  orbits converge to the Hartree-Fock states and this
correction becomes independent of the choice of the single-particle basis.
Note that the prescription of Eq.~(\ref{eq:monopole}) modifies only a few crucial
matrix elements while about six millions of them are defined in the
$N_{max}=9$ model space.
%Still the results reported in this work should be associated with this
%modified Hamiltonian and are not predictions of the sole N$^3$LO model.

We also note that the present truncation of the model space, in terms of the number of 
oscillator shells, does not separate exactly the center of mass motion.
Coupled-cluster calculations have shown that the error introduced by this truncation
becomes negligibly small for large model spaces such as the ones employed here and therefore
it does not represent a major issue~\cite{gour2006,Hag.08}.
In calculations of binding energies, it customary to subtract the operator for the kinetic energy of the
center of mass directly from the Hamiltonian. This term automatically corrects for the zero point motion
in oscillator basis but it depends explicitly on the number of particles.
 In this work, we are interested in transitions to states with different numbers of nucleons [(A$\pm$1]
and aim at computing directly the differences between the total energies. Therefore, the above correction should
{\em not} be employed in the present case.
 One may note that the separation of the center-of-mass motion is an issue related to the choice
made for the model space, rather than the many-body method itself. For example, expressing
the propagators directly in momentum space would allow an exact separation. In this situation, 
the transformation between the center-of-mass and laboratory frames for systems with 
a nucleon plus a $A$ nucleons (or 
$(A-1)$ nucleons) core would also be simple.

\subsection{Choice of $\kappa_M$}
\label{calc_kdep}

%%%%%%%%%%%%%%%%%%%%%%%%%%%%%%%%%%%%%%%%%%%%%%%%%%%%%%%%%%%%%%%%%%%%%%%%%
\begin{figure}[t]
\includegraphics[width=1.0\columnwidth,clip=true]{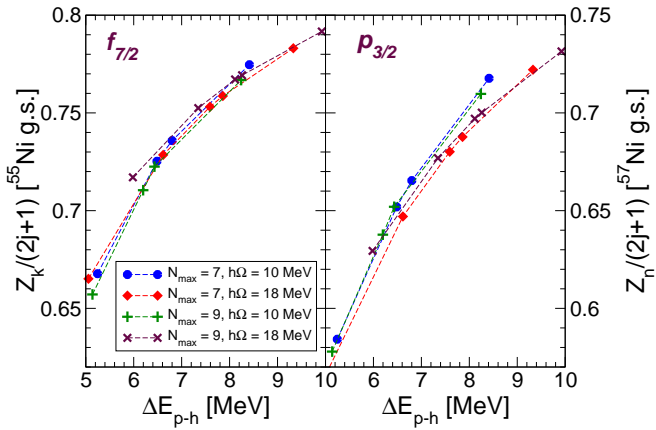}
\caption{(Color online)  Dependence of neutron spectroscopic factors (given as
     a fraction of the independent-particle model value) for the $1p_{3/2}$
     and the $0f_{7/2}$ valence orbits with respect the ph gap $\Delta E_{ph}$.
     For each model space, different points correspond to different choices
     of $\kappa_M$ in the range $0.4-0.7$  MeV.}
\label{sf_vs_k}
\end{figure}
%%%%%%%%%%%%%%%%%%%%%%%%%%%%%%%%%%%%%%%%%%%%%%%%%%%%%%%%%%%%%%%%%%%%%%%%%
 
Eq.~(\ref{eq:monopole}) introduces a single parameter ($\kappa_M$) in our calculations. 
The reason for this modification is that the spectroscopic factors of the valence orbits
are strongly sensitive to the particle-hole gap. 
 This sensitivity is to be expected since  collective modes in the
 $^{56}$Ni core are dominated by excitations across the Fermi surface. Smaller
gaps imply lower excitation energies and higher probability of admixture with
valence orbits.
In order to extract meaningful predictions for spectroscopic factors
it is therefore necessary to constrain the Fermi gaps for protons and neutrons  
to their experimental values. 

To investigate this dependency we repeated our calculations 
for values of $\kappa_M$ in the range $0.4-0.7$ MeV.
 Fig.~\ref{sf_vs_k} shows the resulting neutron spectroscopic factors for
the valence $p_{3/2}$ quasiparticle and $f_{7/2}$ quasihole.
These are plotted as a function of the calculated particle-hole gap
$\Delta E_{ph}=\varepsilon^+_{1p_{3/2},n=0}-\varepsilon^-_{0f_{7/2},k=0}$.
 The results correspond to model spaces of different dimensions (eight or
ten oscillator shells) and oscillator frequencies ($\hbar\Omega=10$ or 18 MeV).
 The gap $\Delta E_{ph}$ increases with $\kappa_M$ but the 
dependence on the model space is weak. We notice that, once the experimental
value of $\Delta E_{ph}$ is reproduced, the spectroscopic
factors are well defined and found to be converged with respect to
the given model space.

All results reported below were obtained with a fixed value of $\kappa_M=0.57$ MeV.
In the $N_{max}=9$ model space and an  oscillator energy $\hbar\Omega=10$~ MeV,
this choice reproduces the experimental gaps at the Fermi surface for both
protons and neutrons to an error within $70$ keV. 
From Fig.~\ref{sf_vs_k} one infers that the calculated spectroscopic factors
are reliable to within $1-2$\% of the independent-particle model value.

\subsection{Convergence with respect to choice of model space}
\label{calc_mspconv}

%%%%%%%%%%%%%%%%%%%%%%%%%%%%%%%%%%%%%%%%%%%%%%%%%%%%%%%%%%%%%%%%%%%%%%%%%
\begin{figure}[t]
\includegraphics[width=0.95\columnwidth,clip=true]{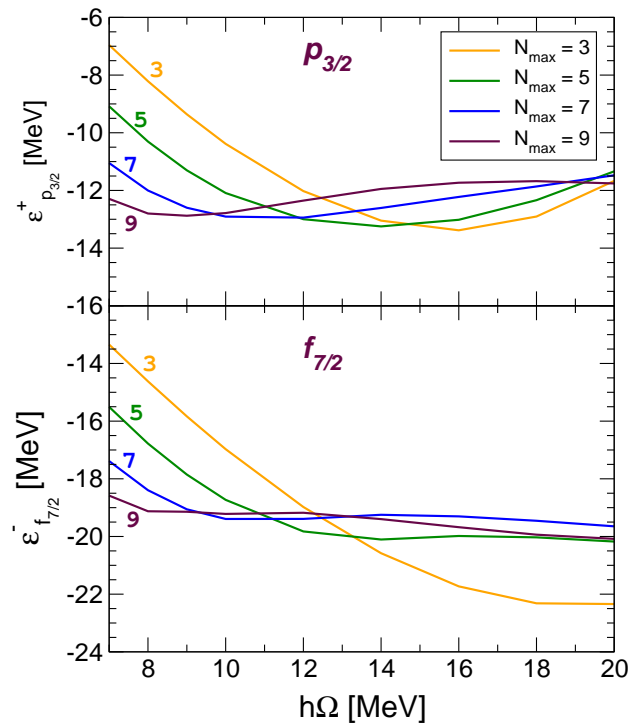}
\caption{(Color online)  Dependence of the neutron $1p_{3/2}$ particle energy
     and the $0f_{7/2}$ hole energy with respect to the oscillator
     frequency and the size of the model space.}
\label{spe_vs_msp}
\end{figure}
%%%%%%%%%%%%%%%%%%%%%%%%%%%%%%%%%%%%%%%%%%%%%%%%%%%%%%%%%%%%%%%%%%%%%%%%%
 
%%%%%%%%%%%%%%%%%%%%%%%%%%%%%%%%%%%%%%%%%%%%%%%%%%%%%%%%%%%%%%%%%%%%%%%%%
\begin{figure}[t]
\includegraphics[width=\columnwidth,clip=true]{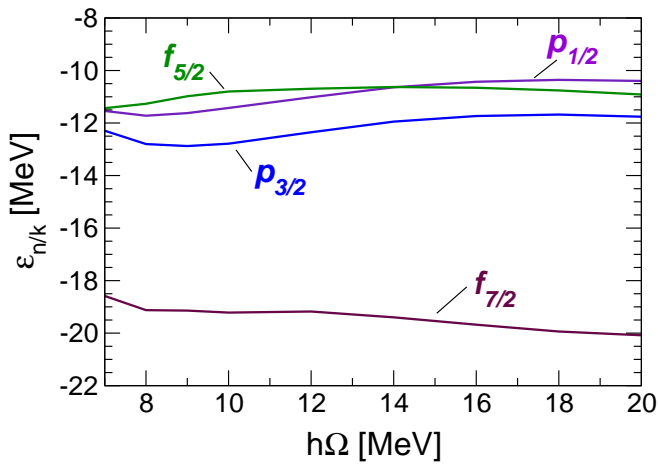}
\caption{(Color online) Dependence of neutron single-particle energies
    on the oscillator frequency. The energies plotted here are the poles
    of $g(\omega)$ corresponding to the valence $1p0f$ orbits.
     Calculations were performed for $N_{max}=9$ and $l\leq 7$.
    }
\label{spe_vs_hw}
\end{figure}
%%%%%%%%%%%%%%%%%%%%%%%%%%%%%%%%%%%%%%%%%%%%%%%%%%%%%%%%%%%%%%%%%%%%%%%%%

Fig.~\ref{spe_vs_msp} shows the dependence 
of the neutron $1p_{3/2}$ particle 
and the $0f_{7/2}$ hole energies with respect to the oscillator
frequency and the size of the model space. 
As can be seen from this figure, the single-particle energies for these two single-particle states
tend to stabilize around eight to ten major shells. 
This finding concords both with coupled-cluster calculations
that employ a $G$-matrix as effective interaction for $^{16}$O, see Refs.~\cite{wloch2005, gour2006},
and with analogous Green's functions studies~\cite{Bar.06}.
It remains however to make an extensive comparison between coupled-cluster theory and the Green's 
function approach in order to find an optimal size of the model space with a given 
nucleon-nucleon interaction.
Finally, we plot  
in Fig.~\ref{spe_vs_hw} 
the neutron valence single-particle energies for all the single-particle states in $1p0f$-shell. 
The latter results were obtained with our largest model space, ten major shells with 
$N_{max}=9$ and the single-particle orbital momentum $l\leq 7$. 
As can be seen from this figure, there is still, although weak,  
a  dependence upon the oscillator parameter.
 To perform calculations beyond ten major shells will require non trivial extensions of our codes.

\section{Results for the spectral function}
\label{res_spctfnct}

%%%%%%%%%%%%%%%%%%%%%%%%%%%%%%%%%%%%%%%%%%%%%%%%%%%%%%%%%%%%%%%%%%%%%%%%%
\begin{table}[t]
\begin{ruledtabular}
\begin{tabular}{lcccccc}
                 & &  \multicolumn{2}{c}{$\varepsilon^+_n$,~$\varepsilon^-_k$}
                 & &  \multicolumn{2}{c}{$Z_n/(2j+1)$,~$Z_k/(2j+1)$}  \\
                 & &  \multicolumn{2}{c}{--------------------------}
                 & &  \multicolumn{2}{c}{--------------------------} \\
                 & &   FRPA   &      Exp.  &  &   FRPA   &  Exp. \\
\hline
$^{57}$Ni: \\
$\nu 1p_{1/2}$    & &  -11.43  &   -9.134   &  &  0.63   &  \\
$\nu 0f_{5/2}$    & &  -10.80  &   -9.478   &  &  0.59   &  \\
$\nu 1p_{3/2}$    & &  -12.78  &   -10.247  &  &  0.65   &  0.58(11) \\
$^{55}$Ni: \\
$\nu 0f_{7/2}$    & &  -19.22  &   -16.641  &  &  0.72   &  \\
 \\
$^{57}$Cu: \\
$\pi 1p_{1/2}$    & &   -1.28  &   +0.417   &  &  0.66   &  \\
$\pi 0f_{5/2}$    & &   -0.58  &            &  &  0.60   &  \\
$\pi 1p_{3/2}$    & &   -2.54  &   -0.695   &  &  0.67   &  \\
$^{55}$Co: \\
$\pi 0f_{7/2}$    & &   -9.08  &   -7.165   &  &  0.73   &  \\
\end{tabular}
\end{ruledtabular}
%\end{center}
 \caption[]{Energies (in MeV) and spectroscopic factors (as a fraction of
    the independent-particle model) for transitions to the $1p0f$ valence orbits, obtained
    for $\hbar\Omega=10$ MeV, $N_{max}=9$ and $\kappa_M=0.57$ MeV.
     The experimental single-particle energies are taken from~\cite{toi.96}.
    The measured spectroscopic factor for transfer between the ground states
    of $^{57}$Ni and $^{56}$Ni is from Ref.~\cite{Yur.06}.
    }
\label{tab:sf}
\end{table}
%%%%%%%%%%%%%%%%%%%%%%%%%%%%%%%%%%%%%%%%%%%%%%%%%%%%%%%%%%%%%%%%%%%%%%%%%

Our results for spectroscopic factors for the $1p0f$-shell valence orbits
and the corresponding single-particle energies are collected 
in Table~\ref{tab:sf}.
In general, the modified N$^3$LO interaction predicts single-particle energies
about $2-3$ MeV lower than the experimental ones.
The Coulomb shift between corresponding neutron and proton orbits is
calculated to be about $10.2$ MeV and it is closer to the empirical value
of $9.5$ MeV.
For the oscillator parameter chosen, $\hbar\Omega=10$ MeV, we obtain an
inversion of the $1p_{1/2}$ and $0f_{5/2}$ excited states in $^{57}$Ni,
with respect to the experiment. However, this discrepancy disappears
for larger values of $\hbar\Omega$~(see Fig.~\ref{spe_vs_hw}).
 This effect is in fact smaller than the residual dependence on the model space
and therefore no conclusion can be made about the ordering for the fully
converged result.
The spectroscopic factor for the transition between the ground states of
$^{57}$Ni and $^{56}$Ni was extracted from high energy knockout
reactions in Ref.~\cite{Yur.06}.
The self-consistent Faddeev random phase approximation 
result for this quantity yield 65\% of the
independent-particle model value, and agrees with the empirical data
within experimental uncertainties.
The theoretical spectroscopic factors for the excited states in $^{57}$Ni
are similar, with the $0f_{5/2}$ state being somewhat smaller, at about 59\%.
A larger value is obtained for knockout to the ground state 
of $^{55}$Ni, which is predicted to be 72\%.
The results for proton transfer to particle (hole) states
in $^{57}$Cu ($^{55}$Co) are only slightly
larger. According to the analysis of Fig.~\ref{sf_vs_k}, it is expected
that these predictions are converged within 1-2\% of the independent-particle model values.

\begin{figure*}
\includegraphics[width=1.6\columnwidth,clip=true]{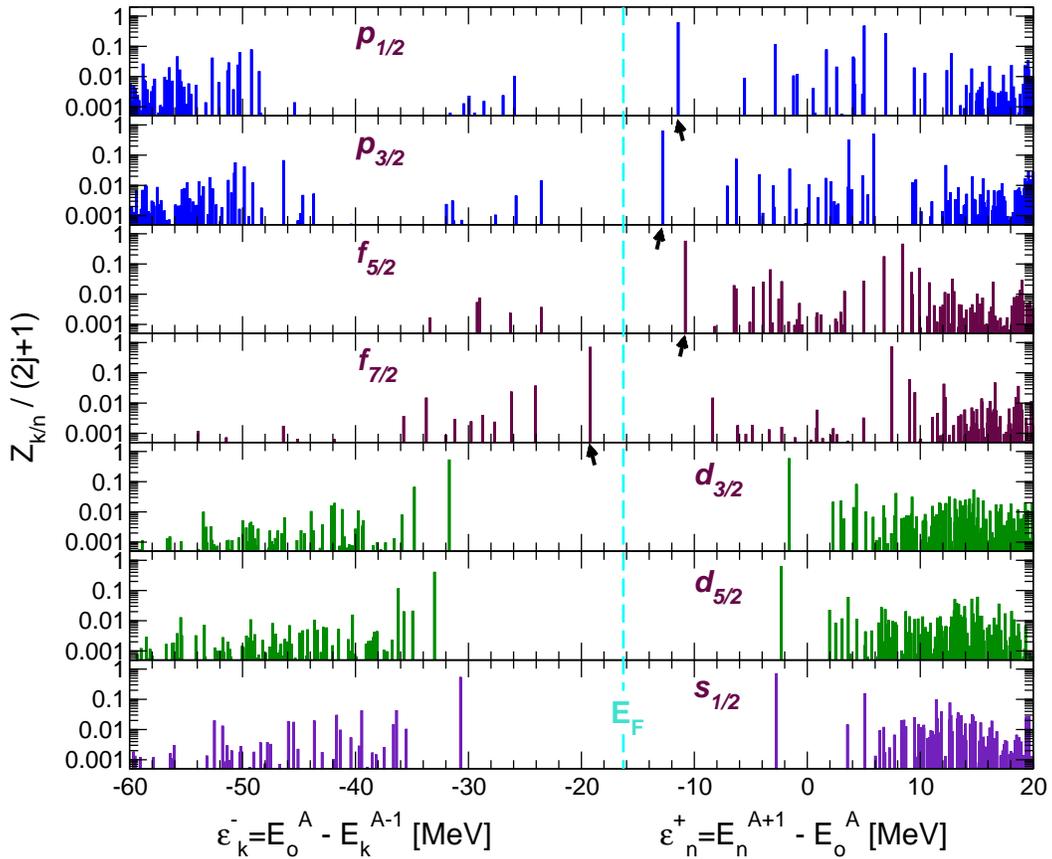}
\caption{(Color online)  Spectral strengths for one-neutron transfer 
     on $^{56}$Ni obtained from the self-consistent single-particle propagator $g(\omega)$.
     Poles above (below) the Fermi energy, $E_F$, correspond to transition to
     eigenstates of $^{57}$Ni~($^{55}$Ni). The respective spectroscopic factors
     are given as a fraction of the independent-particle model value. The quasiparticle poles
     corresponding to the valence orbits of the $1p0f$ shell are indicated
     by arrows. A logarithmic scale was chosen to put in stronger evidence the
     distribution of the fragmented strength.
     Results are for $\hbar\Omega$=10~MeV, $N_{max}$=9 and $\kappa_M$=0.57~MeV.}
\label{fig:SpctFnct}
\end{figure*}

Past studies~\cite{Gou.69,ni56e,ni56f} have questioned whether 
low-energy quasiparticle states in $^{57}$Ni are strongly admixed
to excitations of a soft $^{56}$Ni core.
 The results obtained here do not suggest substantial differences
with respect to other known closed shell nuclei.
The spectroscopic factors from Table~\ref{tab:sf} are in line with 
observations from stable nuclei~\cite{lap93,Gad.08} and 
support the hypothesis that $^{56}$Ni is a good closed-shell nucleus.
In our calculations we find that the $1p_{3/2}$ quasiparticle state of $^{57}$Ni
carries 65\% of the strength for this orbit. Another 20\% is located in the
particle region below $\varepsilon^+_n$=2~MeV (above this energy strength
associated with the $2p1f0h$ shell starts to appear), and about 3\% is in the
hole region above $\varepsilon^-_k=-40$ MeV (see Fig.~\ref{fig:SpctFnct}).
Similarly, the $0f_{7/2}$ state has 72\% of the independent-particle model  
strength in the quasihole
peak (the ground state of $^{55}$Ni), 10\% in the fragmented hole region,  and 
3\% in the fragmented particle region.
This analysis confirms that the main mechanism responsible for the quenching
of the spectroscopic factors lies in the admixture between single-particle 
states and collective excitations in the region of giant resonances~\cite{Dic.04}.
Due to these correlations a large part of the missing strength from the 
valence peak is shifted and spread over an adjacent region about $15-20$ MeV wide.
 Further reduction of the spectroscopic factors comes from the mixing
with configurations at much higher energies and momenta and is accounted
for through the energy dependence of Eq.~(\ref{eq:h_BHF}).

%%%%%%%%%%%%%%%%%%%%%%%%%%%%%%%%%%%%%%%%%%%%%%%%%%%%%%%%%%%%%%%%%%%%%%%%%
%%%%%%%%%%%%%%%%%%%%%%%%%%%%%%%%%%%%%%%%%%%%%%%%%%%%%%%%%%%%%%%%%%%%%%%%%

The information carried by the calculated single-particle propagator $g(\omega)$ is
collected in Fig.~\ref{fig:SpctFnct} for neutrons and partial waves
up to $l$=3. The plot shows the spectral strength associated with each pole
of $g(\omega)$.
%\begin{eqnarray}
% S(\omega) &=& \sum_\alpha \, 2 \pi {\cal Im} g_{\alpha \alpha}(\omega)
%\nonumber  \\
%   &=& \sum_k + \sum_n  \; .
%\end{eqnarray}
Fragments below the Fermi surface ($E_F$) refer to the separation of a neutron
($^{55}$Ni), while those above correspond to neutron addition ($^{57}$Ni).
The poles corresponding to the $1p0f$ valence orbits are indicated by 
arrows. % in Fig.~\ref{fig:SpctFnct}.
These are the same single-particle energies that have been discussed
in Sec.~\ref{calc_mspconv} and Tab.~\ref{tab:sf}.
As noted above, the  fragments found at slightly higher energies (just
above $\varepsilon^+_n\sim -8$ MeV) originate from the mixing of these orbits with
two-particle-one-hole configurations and collective excitations of the nucleus.
 The overall fragmentation effect is substantial but not strong enough to destroy
the single-particle character of the principal quasiparticle peaks (note that a
logarithmic scale has been chosen in Fig.~\ref{fig:SpctFnct} in order to make smaller
fragments of the spectral distribution more visible).
 This observation supports the use of valence single-particle states as the relevant
degrees of freedom that govern low-lying excitations, as assumed
in conventional shell-model applications.
We note that the question whether a system can be approximated as a good shell closure is better
addressed by analyzing the spectroscopic factors and strength distribution rather
than occupation numbers, since the latter  
are integrated quantities~\footnote{Occupation numbers are normally defined in terms
of the density matrix, which involves an integral sum over each hole pole of
Eq.~(\ref{eq:g1}). Especially for deeply bound orbits, it is possible that a strong
fragmentation pattern still leads to large occupation numbers.}.
While unoccupied states can be probed by the addition of a nucleon, occupied states
are accessed by knockout to states of the $A-1$ nucleon system. A similar fragmentation
pattern is therefore seen for the $0f_{7/2}$ orbit but reversed below the Fermi surface.
Interestingly enough, the Faddeev random phase approximation 
predicts that states corresponding to orbits in the $1s0d$
and $2s1d0g$ shells maintain a strong single-particle character even though they are further
apart from the Fermi surface. 
The fragmentation of these orbits requires excitations across shells of different
parity (e.g. $1s0d$ and $1p0f$) and could become stronger if the energy difference among
major shells is reduced.
Indeed, a comparison of our results with electron scattering measurements on
$^{58}$Ni~\cite{Mou.76} suggests that the N$^3$LO interaction tends to overestimate
the gaps between these major shells.
Note that in the present calculations the $2s1d0g$ quasiparticles are found at energies
of about $-3$ MeV and overlap with the fragmented $1p0f$ states.

 Far from the Fermi energy $E_F$, the mixing with complex configurations becomes strong and
it is no longer possible to identify sharp quasiparticle and quasihole states. Still, the energy
region occupied by the major shells can be identified clearly. The N$^3$LO interaction places the states associated with the $1s0d$ shell between $-60$ and $-30$~
MeV, while the $0p$-shell states appear below $-50$~MeV.
 Other hole fragments are observed around $-30$~MeV for the $1p_{1/2}$, $1p_{3/2}$
and $0f_{5/2}$ partial waves. These originate from particle states in the $1p0f$ shell
that are partially occupied due to the smearing of the Fermi surface. Nucleon knockout
from these orbits requires little energy transfer and leads to low-lying states
in $^{55}$Ni or~$^{55}$Cu.
 These states originate from the mixing of particle orbits with
two-hole-one-particle configurations~($R^{(2h1p)}$) in the Dyson equation. Still, the 
$^{55}$Ni ground state is strongly influenced by the $0f_{7/2}$ hole component.
%%One can rephrase this observation by saying that correlations in the two-hole-one-particle channel
%%admix with particle states and lower them below $E_F$.

Analogous fragmentation patterns extend to the shells further away from the Fermi surface,
although these  are not shown in Fig.~\ref{fig:SpctFnct}. 
On the particle side, $\varepsilon^+_n=0$ MeV
marks the threshold for the single-particle continuum in the $(A+1)$-nucleon system.
Above this, the exact spectral function becomes a continuous function of energy.
In the present calculations a structure of separate poles is found due to the discretization
of the model space. A continuum distribution also develops for the hole part of
the exact spectral function below the energy $\varepsilon^-_k$=$(E_0^A - E_{k'=0}^{A-1}) - S_N^{A-1}$,
where $S_N^{A-1}$ is the one-nucleon separation energy from the ground state of $A-1$ particles.
 The distribution on both sides of the Fermi surface is similar but not fully symmetric,
the strength being stronger at large positive energies. This is because the $(A+1)$-nucleon
system can access a larger phase space than a single hole within the $A$-nucleon ground state.
 This asymmetry is already observed at the level of the self-energy $\Sigma^{\star}(\omega)$,
a result in line with  available fits of global optical potentials~\cite{Cha.06,Cha.07}.

For the proton case, the poles of $g(\omega)$ correspond to the addition~(removal) of a proton to 
the eigenstates of $^{57}$Co~($^{55}$Cu). The corresponding spectral strength is substantially the same
as that discussed for neutrons due to the almost exact isospin symmetry of the nuclear force.
However, it is shifted to higher energies by the  Coulomb repulsion.

\section{Conclusions}
\label{concl}

The aim of this work has been to extend large-scale calculations 
of self-consistent Green's functions to medium mass nuclei and to
investigate the properties of the single-particle spectral function
of $^{56}$Ni.

Many-body Green's functions hold a number of interesting mathematical properties.
Since one aims at obtaining excitations relative to a reference nucleus
calculations scale more gently when increasing the number of particles
as opposed to direct large-scale diagonalization methods.
Only connected diagrams are summed to all orders so that the extensivity
condition is satisfied \cite{bartlett2007}. Moreover, the self-consistent approach
provides a path to ensure the conservation of basic macroscopic quantities.
However, the greatest advantage of the self-consistent Green's function 
formalism is that its building blocks,
the many-body propagators, contain information on the response to several
particle transfer and excitation processes. Therefore, they can be directly
compared to a large body of experimental data.
%These pieces of information are also linked to one another by the self-consistency approach.
Due to these characteristics the formalism can be used to gain unmatched
insights into the many-body dynamics of quantum mechanical systems.
Within this framework, the Faddeev random phase approximation method proposed
in Ref.~\cite{Bar.01} is a good candidate to pursue {\em ab initio} studies
of medium mass isotopes.

In this work we have presented the basic details of calculating the Faddeev
random phase approximation expansion and discussed 
results for the spectral function of $^{56}$Ni. This is the first application  of the self-consistent 
Green's function approach to  
the $1p0f$ shell region. The calculations employ the  
chiral N$^3$LO two-nucleon interaction, with a modified monopole to account
for missing many-nucleon forces. In addition to  this one-parameter modification
of the Hamiltonian, the only remaining parameters that enter our calculations are those 
defining the nucleon-nucleon interaction.

Calculations were performed in models spaces including up to ten major
oscillator shells. These large spaces are large enough to  allow
for a sophisticated treatment of long-range correlations.
 The quasiparticle and quasihole energies of the $1p0f$ valence orbits were found
to be rather well converged. In the largest calculations they appeared to be
almost constant for oscillator frequencies in the range $\hbar\Omega\in [8,20]$ MeV.
These convergence properties are possible thanks to a prediagonalization
of the effects of short-range correlations. This is done using the $G$-matrix technique 
\cite{Hjo.95,CENS.08}
to resum ladder diagrams outside the model space.
Our results put in evidence the strong sensitivity of spectroscopic factors
on the particle-hole gap at the Fermi surface. For the $N=28$ and $Z=28$ subshell closures
the bare N$^3$LO potential fails in describing the experimental
gap (in an analogous way to other realistic two-nucleon interactions~\cite{caurier2005}).
This effect has been attributed to missing three-nucleon interactions~\cite{Zuk.03}.
It is found that a proper correction of few monopole terms of the Hamiltonian
allows us to extract reliable results for the fragmentation of single-particle strength.

Fully self-consistent Faddeev random phase approximation 
calculations have till now only been presented  
for $^{16}$O. The extension to accurate {\em ab initio}
calculations in the $1p0f$ shell represent a major technical advance.
However, no substantial use of parallel computation has been made
in applying this formalism.
Improvements in numerical algorithms are still possible and it is 
expected that they will allow a better treatment
of fragmentation in the self-consistent approach, as well as pushing the
limits of present calculations well beyond mass $A=56$.
Another obvious extension is the inclusion of explicit three-nucleon 
forces. Within the framework of self-consistent Green's function  
theory this has already been achieved for
nuclear matter studies~\cite{Soma2008}. Similar developments can be
expected for finite nuclei as well.

For open shell systems with weakly bound states and/or resonances, one needs 
a single-particle basis which can handle continuum states, as done in~\cite{Hag.07}.
Normally, this leads to a much larger space and may require parallelization of our codes.
On the other hand, the effective interaction among valence-space quasiparticles
is already generated in the present calculations and can be used for standard
shell-model calculations in one or two major shells.
The issue of degenerate unperturbed states for open shells systems has also
been  addressed for Green's functions theory in Ref.~\cite{Slu.93} by using 
a Bogoliubov-type quasi-particle transformation.
In a self-consistent treatment, one may improve on this approach by extending the
Faddeev-RPA method to include explicit configuration mixing between the nucleons inside
the open shell.

The $N=28$ and $Z=28$ subshell closure has also attracted recent experimental
interest following the discussion of whether the low-lying states
of $^{57}$Cu are strongly fragmented due to a soft $^{56}$Ni
core, see for example Ref.~\cite{ni56f}.
 While no direct experimental information is available for the 
transition between these two isotopes, the spectroscopic factor for neutron
knockout from $^{57}$Ni has been measured in Ref.~\cite{Yur.06}.
The present calculations describe well the quenching of the
experimental cross section.
% and lie within the error bas estimated from the model dependence of the reaction analysis.
 At the same time, we report predictions for both proton and neutron
transfer to the other valence orbits around~$^{56}$Ni. These calculations can thereby provide 
theoretical benchmarks for
the forthcoming experiments of Refs.~\cite{ni56c,ni56d}.
These spectroscopic factors are all in the range
of 60\%-70\% of the independent-particle model value and  
in fair agreement with the observation of valence states
in several stable nuclei~\cite{lap93}.
 The fragmentation pattern of valence orbits predicted by the Faddeev random phase approximation 
is also found in substantial agreement with what is known for closed shell
nuclei~\cite{Dic.04} and supports the description of $^{56}$Ni as a
doubly magic nucleus.
Finally, we note that the effects of admixing configurations with several
particle-hole excitations are not included in the present study. These effects
can be accounted for by using configuration interaction (shell-model) methods.
However, based on the analysis of Refs.~\cite{Hor.07,Gou.08,newpap} these
corrections are not expected to be dominant.

%%%%%%%%%%%%%%%%%%%%%%%%%%%%%%%%%%%%%%%%%%%%%%%%%%%%%%%%%%%%%%%%%%%%%%%%%%
\acknowledgments
One of the authors (C.B.) would like to acknowledge several useful discussions
with W.~H. Dickhoff and D.~Van Neck.
%%%%%%%%%%%%%%%%%%%%%%%%%%%%%%%%%%%%%%%%%%%%%%%%%%%%%%%%%%%%%%%%%%%%%%%%%%

\end{document}